\documentclass[10pt,letterpaper]{article}
\usepackage{opex3}
\usepackage{amsmath}
\usepackage{cite}
\usepackage{lineno}

\begin{document}
\title{Dark spots along slowly scaling chains of plasmonic nanoparticles
}
\author{Gianluigi Zito,$^{1,*}$ Giulia Rusciano,$^{1,2}$ and
Antonio Sasso$^{1,2}$}
\address{$^1$Dipartimento di Fisica E. Pancini, Universit\`a degli studi di Napoli Federico II, \\
Complesso Univesitario Monte S.Angelo, Via Cintia, 80126-I Napoli, Italy
}
\address{$^2$Istituto Nazionale di Ottica, Consiglio Nazionale delle Ricerche, \\
Via Campi Flegrei 34 - 80078 Pozzuoli (NA), Italy
}
\email{$^*$zito@fisica.unina.it}

\begin{abstract*} \noindent We numerically investigate the optical response of slowly scaling linear chains of mismatched silver nanoparticles. Hybridized plasmon chain resonances  manifest unusual local field distributions around the nanoparticles that result from symmetry breaking of the geometry. Importantly, we find localization patterns characterized by bright hot-spots alternated by what we term \textit{dark} spots. A dark spot is associated to dark plasmons that have collinear and antiparallel dipole moments along the chain. As a result, the field amplification in the dark interjunction gap is extinguished for incident polarization parallel to the chain axis. Despite the strong plasmonic coupling, the nanoparticles on the sides of this dark gap experience a dramatic asymmetric field amplification with amplitude gain contrast $> 2 \times 10^2$. Remarkably, also for polarization orthogonal to the axis, gap hot-spots form on resonance. 
 \end{abstract*}
\;



%
%
%
%

\section{Introduction}
\noindent One major success of Plasmonics relies on the extraordinary possibility of electromagnetic field localization at the nanoscale, mediated by collective electron excitations at suitable metal-dielectric interfaces\cite{MaierBook:2007}.
A homodimer of nanoparticles (NPs) with dimensions $ \ll \lambda_{\rm o}$ (free space wavelength) provides the basic element of most plasmonic architectures. The local photonic density of states can be amplified of several orders of magnitude into volumes comparable with molecular length scales into so-called hot-spots\cite{Halas:2011}. 
Near- and far-field responses supported by the dielectric geometry of the system are determined by the hybridized localized surface plasmon-polariton (LSP) modes\cite{Nordlander:2004}. Engineering of such optical responses has provided a formidable variety of nanophotonic tools that include nanoantenna devices\cite{Giannini:2011,Brown:2010}, surface-enhanced Raman scattering (SERS) substrates\cite{Zito:2015,Cecchini:2012,DeRosa:2015,Isticato:2013,Matteini:2015}, integrated optical devices\cite{Fang:2015}, metasurfaces\cite{Lin:2013}, plasmon-induced loss devices\cite{Ndukaife:2015}, etc.
In general, when the symmetry of a plasmonic configuration of nanoparticles is broken, a larger variety of coupling mechanisms between NP plasmons are possible\cite{Brown:2010}, which exponentially broadens the capability of light interaction and manipulation. \\
\indent Recently, Chen \textit{et at.} reported on the possibility to light specific hot-spots into NP aggregates on a metal film \cite{Chen:2015}. In fact, the coupling between NPs and film support charge density modes that show special charge accumulations over the metal-dielectric boundaries and determine specific near-field distributions. The possibility to excite special hot-spots therefore arises from the symmetry breaking introduced by the metal film.\\
\indent  In this study, we investigate the optical response of a linear chain of six mismatched silver nanoparticles (linear hexamer) under plane wave illumination. The chain modes \cite{Tserkezis:2014} of our linear hexamer provide a rich variety of hot-spot patterns in the gaps between the various NPs, deterministically controlled by the excitation energy. The novelty is that for special antibonding modes, strongly coupled collinear dark plasmons with internal, antiparallel dipole moments can totally quench specific hot-spots, damping the scattered near-field down to the level of the background field. In contrast to cold spots of isolated nanoparticles\cite{Haggui:2012}, we term such an anomalous gap as \textit{dark} spot. Remarkably, we find an exceptional asymmetric near-field distribution around the NPs nearest to the dark gap.  The amplitude-gain contrast is up to $\sim$ 228. To the best of our knowledge, this possibility has never been predicted before. \\

\section{Slowly Scaling Linear Chain of Nanoparticles}

Chains of mismatched NPs have been used to increase near-field enhancement in plasmonic applications such as SERS spectroscopy\cite{Ding:2010,Tserkezis:2014} and near-field chemical scanning\cite{Hoppener:2012,Rusciano:2014}. Indeed, chains of multiscale NPs may enable huge enhancement of the local electric field\cite{Li:2003}. In this work, we study a (relative) simpler system with slow scale variation. We calculate its optical response as a function of the overall length scale introducing an isotropic scale factor $f$ (global parameter). The starting geometry consists of six nanoparticles with radii, from the biggest to the smallest, $R_{\{1,2,3,4,5,6\}} = \{13, 10, 7, 5, 4, 3\}$ nm, in this order, and with gaps measured between closer points $g_{\{12,23,34,45,56\}} =\{ 1.5, 1, 1, 1, 1\}$ nm. For this particular geometry, where only $g_{12} = 1.5$ nm, we empirically found an interesting case of LSP modal interference. In Fig. 1(a), the geometry is depicted for $f = 1.0$. \\
\indent We carried out 3$D$ numerical simulations solving the full electrodynamic problem with a finite element method-based commercial software (Comsol 4.3b).  We considered spherical nanoparticles of silver with dielectric function from Johnson and Christy\cite{Johnson:1972} embedded in a surrounding vacuum medium. The spherical simulation region was embedded into a perfectly matched layer with outer scattering boundary conditions\cite{Zito:2015}. Minimum mesh element was 0.07 nm. The simulations were carried out with wavelength step of 2.5 nm (0.2 nm for finer inspections). Relative error tolerance was set to 1$\times$10$^{-7}$. With a quad-core processor Intel i5-3570K @3.4GHz and 32 GB of RAM, simulation time was ca. two minutes per wavelength. The dipole moments of the NPs associated to the various chain modes were calculated integrating the polarization density vector over the NP volumes and, for control, by the relation $\vec{p}(\vec{r}) =\int_{\partial V} \sigma(\vec{r_{0}})(\vec{r_{0}}-\vec{r}) {\rm d^2}\vec{r_{0}}$, where $\sigma$ is the surface charge density given by the relation $\sigma (\vec{r_{0}}) = \epsilon_{0}n^2 \hat{u} \cdot \left[\vec{E_{s}} (\vec{r_{0}} )+ \vec{E_{0}} (\vec{r_{0}})\right] $, whereas $\partial V$ is the surface of the nanoparticles with $\hat{u}$ normal versor to the boundaries and $n$ the refractive index. The scattering far field response was calculated by using an auxiliary surface enclosing radiating NPs.\\
\indent The scattering cross section (SCS) spectrum of the linear hexamer excited by a plane wave with polarization parallel to the chain axis is shown in Fig. 1(b). The family of curves is  obtained by varying $f$. The SCS presents multiple peaks across the UV-visible range related to high-order LSP resonances. We will refer to these modes indicating the lower energy radiant mode of interest as RM$_{0}$ and progressively higher order radiant modes as RM$_{1,2,3}$. Of course, the spectral positions of these modes evolve with the system size as clearly visible in Fig. 1(b). An increase of the surrounding $n$ produces a redshift of the peaks without affecting their overall structure (not shown). At RM$_{0}$ the near-field is maximized. It is worth noticing that this mode, spectrally located just on the left of the SCS peak, is not a dipole mode. The dipole mode is located on the right of the SCS maximum and shows inferior near-field enhancement. The spectral dip indicated as subradiant mode (SM) in the SCS is of particular interest to our discussion as will be clear later. \\
\begin{figure*}[t]
	\centering
	\setlength{\unitlength}{\textwidth}
		\includegraphics[width = 13cm]{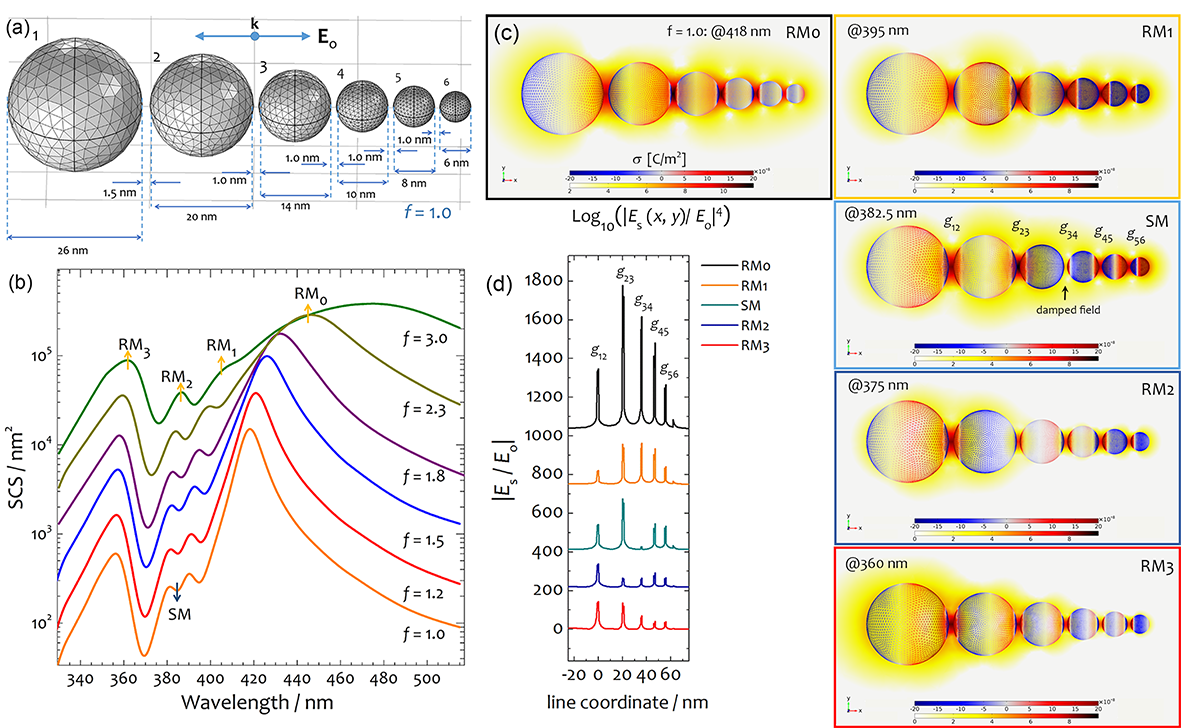}
	\caption{(a) Schematic layout of the hexamer of silver spheres ($f = 1.0$). (b) Family of SCS spectra parametrized by the isotropic scale factor $f$ for plane wave linearly polarized along the axis chain. (c) Surface charge density modes (wireframed on NP surfaces) and overlaid $G$-map for different LSP modes at $f = 1.0$, calculated in the $xy$ midplane of the chain for polarization parallel to the chain axis. (d) Field enhancement cross-section along the chain axis for the same LSP modes depicted in (c) (curves are vertically shifted for clarity).}
	\label{fig:Figure_1}
\end{figure*}
\indent  We will focus first on the case $f = 1.0$. The characteristic surface-charge density modes $\sigma(\lambda_{\rm o})$ on each nanoparticle are plotted in Fig. 1(c)  for RM$_{i}$ $(i = 1, 2, 3)$ and SM. In the heterochain, translational symmetry is broken. As a consequence, the  mode profiles $\sigma (\lambda_{\rm o})$ become characteristic of each NP since determined by the local topology of the system and the interaction with the surrounding NPs.  Overlaid in panels of Fig. 1(c), we also report  the corresponding spatial distributions of the fourth power of scattered field amplitude $E_{\rm s}$, normalized to the incident plane-wave amplitude $E_{\rm o}$, i.e. the approximated SERS enhancement factor $G(x, y, z) = |E_{\rm s}/E_{\rm o}|^4$ (indicated as $G$-map). The five main states of interest extracted in Fig. 1(c) correspond to the spectral positions indicated in the SCS of Fig. 1(b) for $f = 1.0$. The chain modes of the linear hexamer provide a rich variety of hot-spot patterns in the NP gaps. Such patterns consist of sequences of relative maxima of the local field that can be deterministically controlled by varying the wavelength of the excitation radiation.  In fact, the field amplification does not follow a cascade progression towards the smallest NP, which requires a rapidly scaling chain\cite{Li:2003}.  We find that local field amplification $|E_{\rm s} / E_{\rm o}|$ in the various gaps of the chain is highly dependent on the particular LSP. In Fig. 1(d), we plot it along the chain axis for the resonant modes illustrated in Fig. 1(c). The left edge of NP$_2$ [Fig 1(a)] is the origin of the coordinate system. As can be seen, the position of maximum amplification moves from one gap to another following a scheme that can be associated to each LSP mode. Indeed, at a fixed wavelength, the local surface charge associated to the mode can accumulate on specific regions of the NPs, in equilibrium with the whole system (of course, the overall net charge must be null), and give rise to gap hot-spots of different strength. \\
\begin{figure*}[t]
	\centering
	\setlength{\unitlength}{\textwidth}
		\includegraphics[width = 13cm]{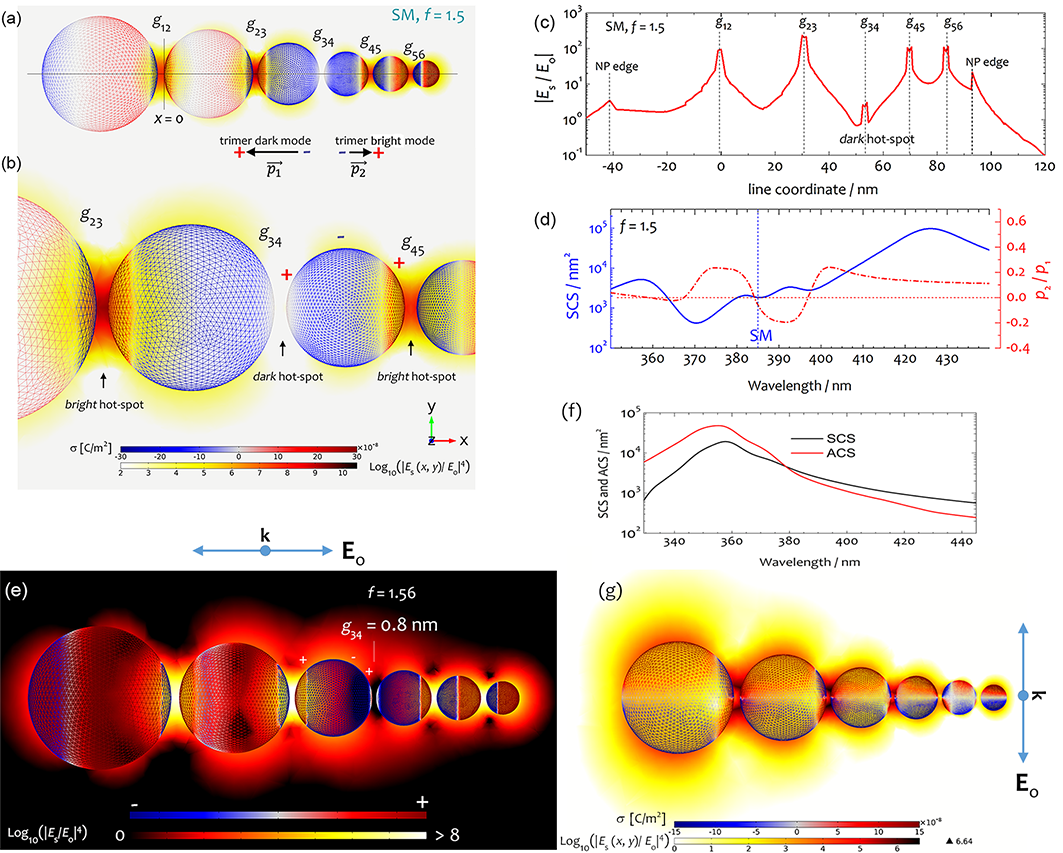}
	\caption{(a) Surface charge density modes and overlaid $G$-map at SM  for $f=1.5$, $\lambda_{\rm o} =$ 390 nm in Fig. 1(b), for polarization parallel to the chain axis.  (b) Close-up around $g_{34}$.  (c) Electric field gain along the chain axis: in $g_{34}$, the gain is damped of nearly 2 orders of magnitude. (d) SCS spectrum (blue) and dipole moment ratio between small trimer ($p_{2}$) and large trimer ($p_{1}$) for polarization along the axis: the dipoles become antiparallel in correspondence of SM. (e) Optimum found at $g_{34} = 0.8$ nm ($f = 1.56$): exactly $E_{{\rm s}} = E_{{\rm 0}}$ in the dark gap (colormap inverted for clarity). (f) Scattering and absorption cross section (ACS) in case of polarization orthogonal to the chain axis, for geometric paramters as in (e). (g) Surface charge density and $G$-map at 365 nm (maximum near-field) for  orthogonal polarization: gap hot-spots form also in this case.}
	\label{fig:Figure_2}
\end{figure*}
\indent Now, let us consider the behavior of the subradiant SM dip that occurs at 382 nm for $f = 1.0$ [Fig. 1(b)]. Inspecting the corresponding mode profile in Fig. 1(c) (SM), we observe a sort of interruption at $g_{\rm 34}$ in the sequence of gap hot spots. In the adjacent gaps, the field is again highly amplified.  The local field is reduced by a factor 20 in $g_{\rm 34}$ with respect to the maximum field gain observed along the chain axis, see SM in Fig. 2(c)-(d).  Looking into this phenomenon, we observe a further decrease of the field amplification for $f = 1.5$. The corresponding mode and field maps are shown in Fig. 2(a)-(b). Indeed, the SM mode further damps the local gain in $g_{34}$ down to $\sim$ 3, while in the adjacent gaps $g_{23}$, $g_{45}$ and  $g_{56}$ the local field is enhanced up to 232-, 110- and 116-fold $E_{\rm o}$, respectively [Fig. 2(c)]. We define the gain contrast as the ratio  $E_{\rm s}(g_{23}) / E_{\rm s}(g_{34})$ between the field enhancements  observed in the adjacent gaps around NP$_{3}$. This ratio evaluates the asymmetry of the amplification on that NP. The maximum contrast of the field gain is therefore equal to a factor $232/3 \simeq 77$, which would correspond to an approximated SERS amplification quenching from $2.9\times 10^9$ to 81, i.e. a decrease of more than seven orders of magnitude ($3.6\times10^7$), followed by a rapid spatial increase to $1.5\times10^8$ in the next gap of the chain [Fig. 2(b)]. A further fine tuning of $f$ allows us increasing the contrast to $229/2 = 114.5$ with $f = 1.56$, always at SM ($\lambda_{\rm o} = 385$ nm). Therefore, when spectrally approaching the SM-dip, the local field gain is nearly turned off in $g_{34}$ and the actual damping depends on the tuning of the modes triggered by the coupling between the NPs. The otherwise bright hot-spot is highly damped. In Fig. 2(c), such a hot-spot has been indicated as dark hot-spot for the reason clarified in the following. \\
\indent To shed light on this phenomenon we calculated the dipole moments of the system as a function of $\lambda_{\rm o}$. After inspecting the surface-charge density $\sigma$ on the NPs in Fig. 2(a), we focused on the partial plasmon modes formed by the first and last three NPs - say large- and small-trimer, respectively. While the first has a weakly radiative symmetry, the second is a bright mode but with smaller dipole moment because of the smaller NP size. Both dipole moments of these trimers, respectively $\vec{p}_{1,2}$ [Fig. 2(a)], are weak at SM.  In fact, $p_{2} \simeq -0.06 p_{1}$ and their sum is $\sim 1/10$ of the hexamer dipole moment induced at the RM$_{0}$, mainly contributed from the weakly-radiative large trimer. We can say that the overall hexamer LSP behaves as a \textit{binary} dark plasmon mode since formed by a dark (large trimer) and a bright mode (small trimer) that have \textit{collinear} antiparallel dipole moments at SM, which decreases the overall dipole moment of the system. Thus the hexamer mode is subradiant. In Fig. 2(d), we can observe the reversal of the polarity of the trimer plasmons in coincidence with the spectral dip SM (blue line). The spectral dip reflects the minimum of total dipole moment of the hexamer.  Therefore, the  opposite dipole moments, induced on average on both trimer plasmons, produce a minimal charge density on the NP surfaces forming the gap $g_{34}$ (nearly neutral). This produces an apparent decoupling between the trimers despite their close proximity ($g_{34} \sim 1$ nm) responsible for the damping of the field in $g_{34}$. This mechanism is therefore at the origin of what we term dark hot-spot in the interjunction between the trimers [Fig. 2(b)-(c)]. Although being a relative maximum of near-field, it is surprisingly weak if compared to the bright hot-spots that the very same NPs on the side of $g_{34}$ create with the next NPs of the chain. \\
\indent  At this stage, we explored the influence of the coupling also varying the size of the gap. Fixed $f = 1.56$, we explored the range $g_{34} =$ 0.6$ - $2.0 nm leaving unchanged the other geometric parameters of the chain. Please note that since $f = 1.56$ all parameters rescale, so that $g_{34} = $ 0.936$-$3.12 nm in the rescaled structure. We find that the contrast reaches a maximum for $g_{34} = 0.8$ nm (i.e. 1.248 nm after rescaling). In fact, as shown in Fig. 2(e), it is possible to have a scattered field exactly equal to the background field, $|E_{\rm s} / E_{\rm o}| = 1$. In this case, the maximum contrast (near field enhancement $g_{23} : g_{34}$) becomes equal to $228/1$ (i.e., $2.7 \times 10^9$ in the SERS enhancement factor). Since the field enhancement is now totally quenched, the dark hot-spot more properly may be indicated as dark spot [Fig. 2(e)]. Changing the value of $g_{12}$ from 1.5 nm reduces the optimal quenching (we explored the range 1.2 - 1.7 nm). \\
 \indent In case of polarization perpendicular  to the chain axis, shown in Fig. 2(f)-(g), the coupling between adjacent NPs gives rise to multipolar splitting of the charge density close to the gaps. As a consequence, also in this case, gap-hot spots can form. Therefore, it is not possible to quench the near-field in the gaps simply rotating the incident polarization. These hot-spots are characterized by two distinct, symmetric maxima in each gap. A $G_{{\rm max}} = 4.4 \times 10^6$ is found at 365 nm (wavelength of maximum near-field) [Fig. 2(g)]. 
  
\section*{Conclusion}

We have theoretically investigated the optical response of mismatched silver nanoparticles forming a slowly scaling linear chain. Our study sheds light on novel phenomena that may arise into  plasmonic systems. Patterns of local field gain can be deterministically excited in the gaps of the chain. The local field gain can be totally quenched in association to a dark chain plasmon with internal antiparallel dipole components along the chain, despite the close proximity ($\sim$ 1 nm) between NPs. In principle, this effect could be used to manipulate the emission properties of quantum emitters in close proximity to the bright and dark gaps. Addressable spatial control of field localization might be employed for novel nanophotonic devices. These results have potential impact for plasmon-enhanced spectroscopies, nanosensing and plasmon-induced loss or plasmonic-force devices. \\
 \indent We acknowledge financial support from Italian Ministry of Education, University and Research, Grant No. FIRB 2012-RBFR12WAPY; and in part from University of Naples Federico II, Compagnia di San Paolo e Istituto Banco di Napoli - Fondazione (LARA).

\end{document}